\documentclass[12pt]{article}
\usepackage{graphicx}

\usepackage{fullpage}
\usepackage{times}
\usepackage{ifthen}

\newcommand{\spacing}[1]{\renewcommand{\baselinestretch}{#1}\large\normalsize}

\title{Binary Stars as the key to understanding Planetary Nebulae}

\author{David Jones$^{1,2}$ \& Henri M.J. Boffin$^3$}
\usepackage{color}

\newenvironment{affiliations}{%
    \setcounter{enumi}{1}%
    \setlength{\parindent}{0in}%
    \slshape\sloppy%
    \begin{list}{\upshape$^{\arabic{enumi}}$}{%
        \usecounter{enumi}%
        \setlength{\leftmargin}{0in}%
        \setlength{\topsep}{0in}%
        \setlength{\labelsep}{0in}%
        \setlength{\labelwidth}{0in}%
        \setlength{\listparindent}{0in}%
        \setlength{\itemsep}{0ex}%
        \setlength{\parsep}{0in}%
        }
    }{\end{list}\par\vspace{12pt}}

\renewenvironment{abstract}{%
    \setlength{\parindent}{0in}%
    \setlength{\parskip}{0in}%
    \bfseries%
    }{\par\vspace{-6pt}}
    
\makeatletter
\def \@maketitle{%
  \newpage\spacing{1}\setlength{\parskip}{12pt}%
    {\Large\bfseries\noindent\sloppy \textsf{\@title} \par}%
    {\noindent\sloppy \@author}%
}
\makeatother

\def\@cite#1#2{$^{\mbox{\scriptsize #1\if@tempswa , #2\fi}}$}

\begin{document}

\maketitle

\begin{affiliations}
\item Instituto de Astrof\'isica de Canarias, E-38205 La Laguna, Tenerife, Spain
\item Departamento de Astrof\'isica, Universidad de La Laguna, E-38206 La Laguna, Tenerife, Spain
\item European Southern Observatory, Karl Schwarzschild Strasse 2, 85748 Garching, Germany
\end{affiliations}

\begin{abstract}
Planetary nebulae are traditionally considered to represent the final evolutionary stage of \underline{\emph{all}} intermediate-mass stars ($\sim$ 0.7 -- 8M $_\odot$).  Recent evidence seems to contradict this picture.  In particular, since the launch of the Hubble Space Telescope it has become clear that planetary nebulae display a wide range of striking morphologies which cannot be understood in a single star scenario, instead pointing towards a binary evolution in a majority of systems.  Here, we summarise our current understanding of the importance of binarity in the formation and shaping of planetary nebulae, as well as the surprises that recent observational studies have revealed with respect to our understanding of binary evolution in general. These advances have critical implications, including for the understanding of mass transfer processes in binary stars -- particularly the all-important but ever-so poorly understood `common envelope phase' -- as well as the formation of cosmologically important type Ia supernovae. 
\end{abstract}

Planetary nebulae (PNe; singular, PN) are the glowing shells of gas and dust observed around stars that have recently left the asymptotic giant branch (AGB) and are evolving towards the white dwarf stage.  The misnomer `planetary nebula' was coined by Frederick William Herschel in the late 18th century, in reference to their resemblance to his recent discovery, the planet Uranus.  
PNe are critical to understanding in many areas of modern astrophysics. They allow us to tackle issues in stellar evolution, stellar populations, gas dynamics, and the formation of dust and molecules, including fullerenes  \cite{kwitter14}. Given their brightness, they are also used to study chemical and dynamical evolution of galaxies \cite{coccato13,magrini16,buzzoni06}, as well as to probe the intracluster medium \cite{gerhard07,arnaboldi03}. Extragalactic PNe are moreover used as standard candles \cite{ciardullo89,ciardullo02}. 

\begin{figure}
\caption{A selection of planetary nebulae known to host binary central stars, highlighting the wide array of morphologies observed in these objects. a) Fleming 1,  b) NGC~5189, c) Shapley 1, d) NGC~6326,  e) The Necklace, f) Henize~2-428, g) Abell 65, h) NGC~1514, i) ETHOS~1, and j) Henize 2-39. Panels reproduced with permission from: ESO \cite{boffin12b} (a); NASA, ESA, and the Hubble Heritage Team (STScI/AURA) (b,e); ESO (c,f); ESA/Hubble and NASA (d); Don Goldman (g); NASA/JPL-Caltech/UCLA, \cite{ressler10}, AAS/IOP (h); \cite{miszalski11b}, Oxford Univ. Press (i); \cite{miszalski13a}, Oxford Univ. Press (j).}
\label{fig:images}
\includegraphics[width=\textwidth]{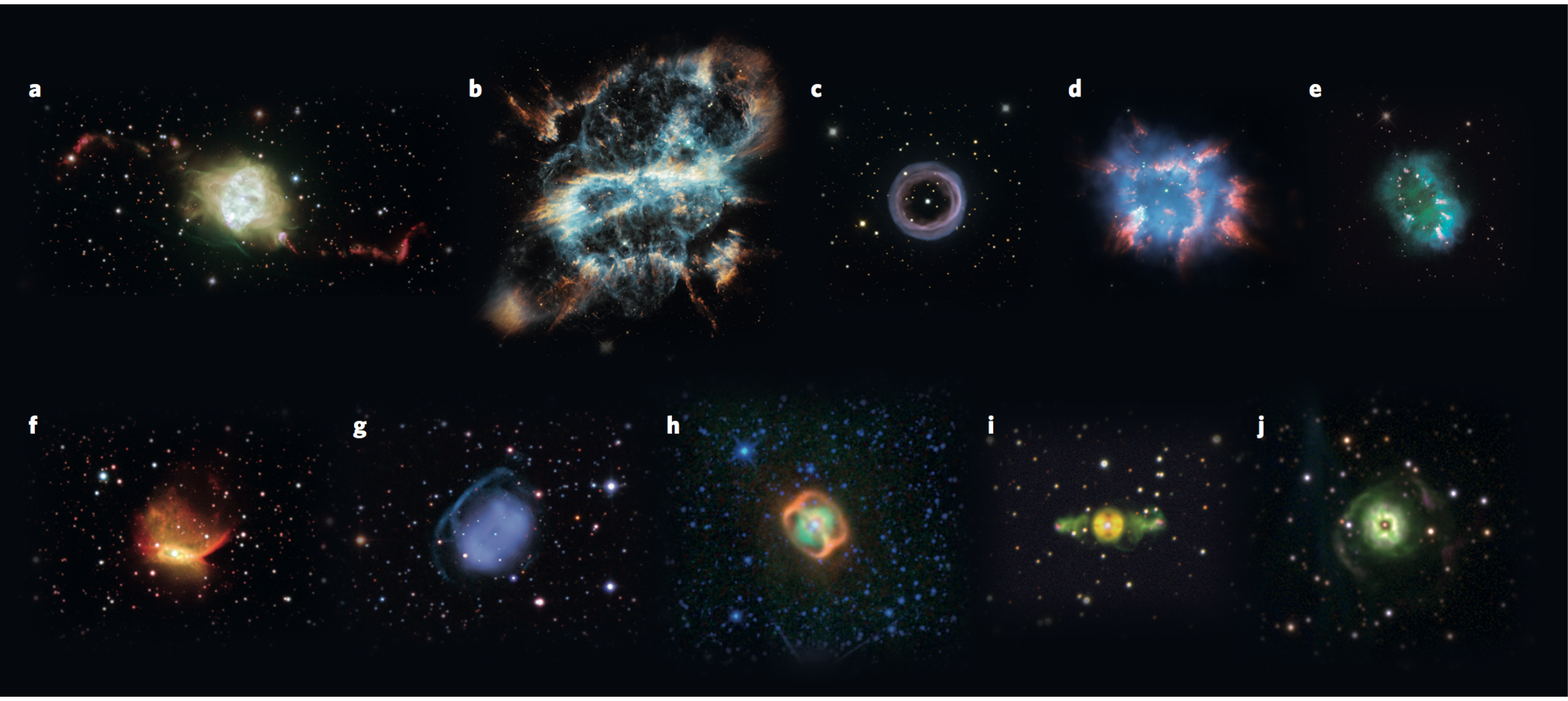}
\end{figure}

Since the late 1970s, PNe have been thought to form as a result of a snow-plough-like mechanism, in which the mass lost by the central star while it is a red giant (as part of a slow, dense stellar wind) is later swept up by a fast but more tenuous wind originating from the now exposed, pre-white-dwarf stellar core\cite{kwok78}.  With some adaptations (namely aspherical winds \cite{kahn85}), this model -- known as the Generalised Interacting Stellar Winds model -- successfully reproduces a broad-range of observed PN properties.  However, when the Hubble Space Telescope allowed these objects to be observed in greater detail than ever before, it revealed both large- and small-scale structures that cannot be explained by this simple model (Fig.~\ref{fig:images}).

\section*{Aspherical morphologies of PNe}
Several contending theories emerged to try and explain the many aspherical morphologies found in PNe (roughly 20\% are found to be spherical, with the rest showing appreciable, and often extreme, deviation from sphericity \cite{parker06}).  The following three received particularly significant attention from the community.
 
 \begin{itemize}
\item \textbf{Stellar rotation.} Rapid rotation can naturally lead to equatorially enhanced mass-loss, which varies with evolutionary phase.  In the case of massive stars, the interactions between different phases of rotationally driven mass loss have been shown to be able to produce highly axisymmetric structures like rings and polar caps \cite{chita08}.  However, in PNe, it has been demonstrated that although rapid rotation may play a role in the formation of nebulae with small deviations from spherical symmetry, it cannot reproduce the most axisymmetic bipolar structures, because this would require rotation rates that are impossibly high for single stars \cite{garcia-segura14}. 

\item \textbf{Magnetic fields.} Strong magnetic fields can act to constrain outflows into a range of collimated structures, both along the rotation axis, in the form of jets \cite{garcia-segura97}, and around the magnetic equator, forming a torus-like feature \cite{matt06}.  However, it has been shown that in isolated PN progenitors, the magnetic fields are neither strong enough or long-lived enough to produce anything but weak deviations from spherical symmetry, instead requiring a binary companion to provide the necessary angular momentum for magnetic-field-driven shaping \cite{nordhaus07,garcia-segura14}.

\item \textbf{Central star binarity.}  The possibility of close-binary PN nuclei was being hypothesized upon as early as the 1970s\cite{paczynski76}, with them being considered key evidence for common-envelope evolution (see below for an explanation).  Since then, binary interactions have become the preferred scenario for the formation of aspherical PNe.  Close binaries offer a particularly clear route to the formation of axisymmetrical structures, in which the mass-transfer processes  
act to deposit more material in the orbital plane of the binary, thus constricting any later winds or outflows into a bipolar shape \cite{nordhaus06}.
Moreover, as previously mentioned, binary interactions can result in a dynamo effect which helps to sustain relatively strong magnetic fields and therefore produce magnetically driven outflows like jets \cite{tocknell14}.
\end{itemize}

\section*{PNe as the result of binarity}
Binary evolution, even when on the main sequence, can be dramatically different to single-star evolution \cite{eggleton11,hurley02,demarco17}. Thus, given the high binary fraction among solar-type main sequence stars \cite{raghavan10}, it is therefore perhaps no surprise that binarity might play an important role in the formation and evolution of PNe.

Theoretically, a link between binarity and nebular structure is expected for several reasons. First, models predict that some PNe should be the outcome of a dynamical event inside a binary system, known as the common envelope (CE)\cite{demarco09}. An AGB star in a binary system will, unless the system is really very wide, interact in some way with its companion. In particular, if the orbital period is short enough, the AGB star will overflow its Roche lobe (the smallest surface of gravitational equipotential that encloses the binary system), which will generally lead to a CE and a spiral-in of the companion\cite{ivanova13}. The final orbit will be very tight, with an orbital period in the range of a few hours to a few days. In this case, the ejected CE is the nebula that will later be ionized by the resulting white dwarf. Such PNe provide us with the best testing ground of the CE interaction \cite{iaconi17}. This is because binaries in these PNe would be `freshly baked': given the short lifetime of a PN (approximately 50\,000 years) and the extremely rapid timescale of the CE itself (generally thought to be a few years or even less\cite{ivanova13}), the CE must have been ejected extremely recently and thus the binaries' orbital properties represent the direct outcome of the CE process. By determining the correlations between the stellar, binary and PN parameters in these objects, one can obtain invaluable constraints on the CE process\cite{hillwig16}. This is particularly useful because the CE process is unfortunately one of the least well understood stages of binary evolution \cite{demarco11}, despite its critical importance in many areas of astrophysics, including in understanding the formation of Type Ia supernovae \cite{ruiter09}. The CE phase is also crucial for understanding classes of objects such as cataclysmic variables, novae, low-mass X-ray binaries, and symbiotic stars. Moreover, the CE is a primary issue in the theory of formation of binary compact objects that give rise to detectable gravitational waves. 

Furthermore, mass transfer in a detached (wide) system can also lead to density enhancements in the orbital plane, and lead to aspherical PNe\cite{boffin15}, given the huge mass loss through the AGB's wind and the fact that the wind velocity is comparable to the orbital velocity. Hydrodynamic simulations have clearly demonstrated that the flow structure of a mass-losing AGB star in a wide binary system is very different from the simple, symmetric Bondi-Hoyle accretion flow, with the presence of a large spiral\cite{theuns96,mohamed12}. Such spirals have since then been detected observationally by the Hubble Space Telescope in the carbon star AFGL 3068 \cite{mauron06,kim12} and with the Atacama Large Millimeter/submillimeter Array (ALMA) in R Scl \cite{maercker12}, and in some cases used to determine parameters of the binary producing the aforementioned spiral structure \cite{kim15,kim17}.  Once the AGB star has evolved towards the white dwarf phase, and a PN emerges, one may expect an aspherical nebula to appear \cite{kim12b}. Thus, some PNe would also be expected to contain long-period binaries.

\section*{Observational evidence for the importance of binaries}

\begin{figure}
\caption{The PN Abell 63, the central star of which was the first to be confirmed as a binary (UU Sagittae).  Image based on archival data taken in the light of H$\alpha$+[N~\textsc{ii}] using the Wide Field Camera instrument of the Isaac Newton Telescope.}
\label{fig:A63}
\includegraphics[width=\textwidth]{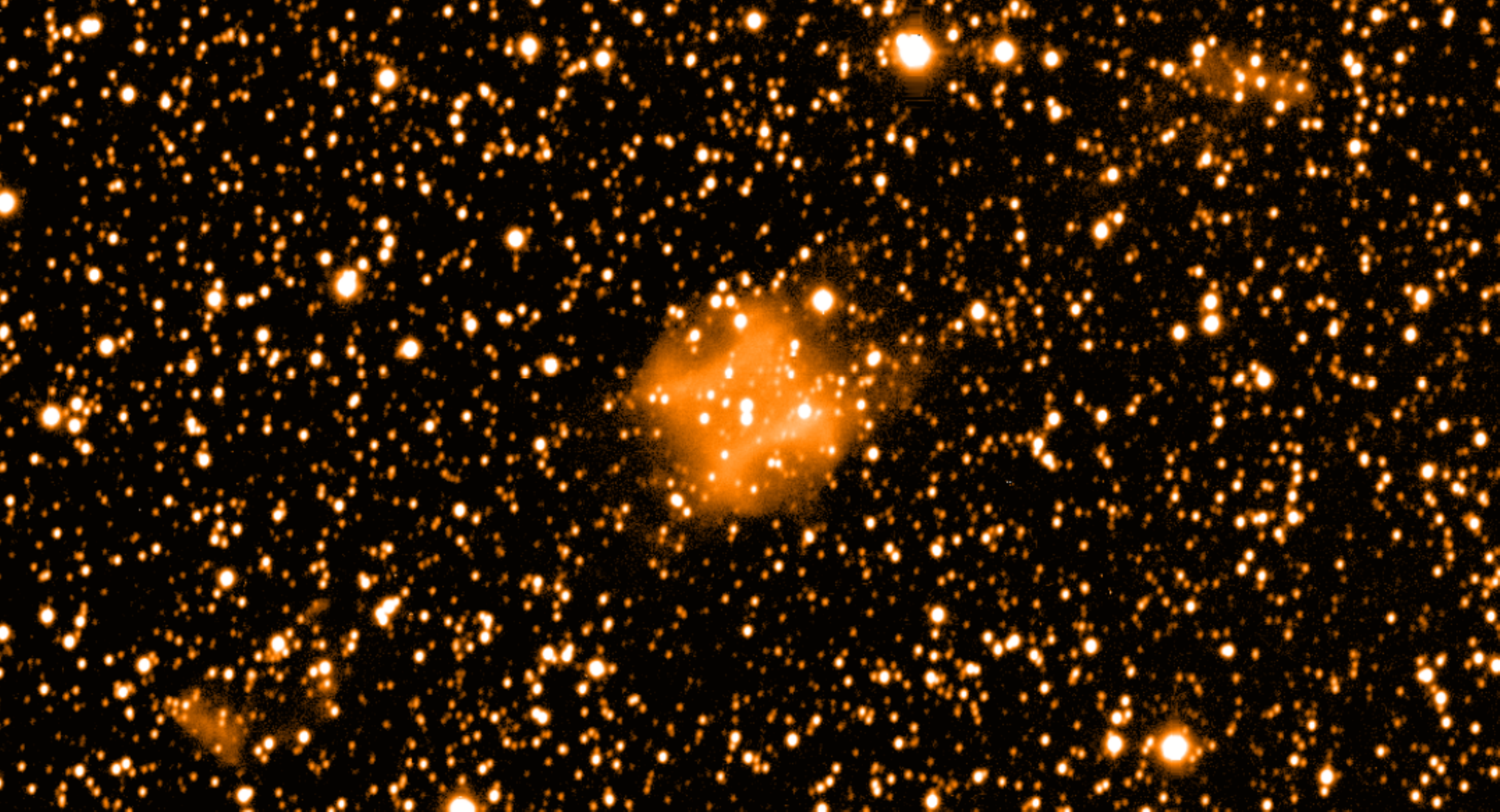}
\end{figure}

The first confirmed binary central star -- UU Sagittae, the central star of Abell~63 -- was found in 1976 through the comparison of the Catalogue of Galactic Planetary Nebulae and the General Catalogue of Variable Stars \cite{bond76}.  Later observations have since shown Abell~63 to be somewhat of an archetype system, displaying many of the features now considered symptomatic of a binary star's evolution (Fig.\ \ref{fig:A63}).  

UU Sagittae is an eclipsing system, showing deep ($\sim$4 mag) eclipses of the primary star (`primary') as well as shallower ($\sim$0.2 mag) eclipses of the secondary star (`secondary'), with a period of 0.465 days \cite{bell94}.  Beyond the variability due to eclipses, the object's light curve shows smooth, sinusoidal variability (amplitude$\sim$0.2 mag) due to what is known as a reflection or irradiation effect \cite{hilditch96}. In such a system, a low-temperature companion is irradiated by the hot primary, resulting in a dramatic increase in temperature and brightness of the companion's face directed towards the primary.  The orbital motion of the binary then results in a differing projection of this heated face, which produces a sinusoidal modulation in the light curve.  Detailed modelling of the system had shown that, in addition to being strongly irradiated by the primary, the companion in this system inflated to approximately twice the radius that would be expected for typical main sequence star of the same mass \cite{afsar08}.  

The nebula Abell~63 was found to consist of a central barrel-shaped structure \cite{pollacco97}, where the symmetry axis lies perpendicular to the binary orbital plane (just as predicted), with more extended, high velocity outflows which were formed a few thousand years prior to the central nebula \cite{mitchell07a}.  Additionally, it has been shown that the relative abundances of the chemical species in the nebular shell show a large discrepancy depending on whether they are calculated using bright collisionally-excited lines or the fainter recombination lines.  This well-known problem is present in almost all astrophysical nebulae for which the ratio of abundances from recombination lines to collisionally-excited lines is usually of order 1--3  \cite{wesson05,garcia-rojas07}. However, in Abell 63 the abundance discrepancy factor (\emph{adf}) is found to be around 10 \cite{corradi15}.  

Since 1976, many more PNe with binary central stars have been discovered, many of which show remarkably similar properties to Abell 63 and its central binary star UU Sagittae \cite{bond90,exter03,exter05,bond00}.  The greatest jump in the known number of close binary stars came from the Optical Gravitational Lensing Experiment (OGLE) survey, a photometric survey designed primarily to probe dark matter using gravitational microlensing events.  Fortunately, the OGLE fields included a number of PNe for which the cadence and sensitivity of the observations allowed their central stars could be checked for photometric variability \cite{miszalski09a}.  This work more than doubled the number of known binary central stars at the time (adding $\sim$20 new systems) thereby allowing for a comparison of the host nebulae to the general population of PNe. It was found that PNe with detectable close binary stars tend to show bipolar morphologies with equatorial rings and extended jet-like features, as well as low-ionisation filamentary structures (Fig.\ \ref{fig:images}) \cite{miszalski09b}.  These morphological features have since been used, with great success, as pre-selectors for targeted surveys for central star binarity \cite{miszalski11a,miszalski11b,corradi11,boffin12b,jones14a,jones15}.

\begin{figure}
\caption{Amplitude of irradiation effect variability as a function of period/separation, inclination and secondary-star spectral type \cite{astroquant} for a base system of a 100,000K, 0.6 M$_\odot$ remnant (taken from evolutionary tracks \cite{millerbertolami16}) in a 1-day orbit inclined at 70$^\circ$.  The horizontal line shows the approximate detection limit of the OGLE survey.  Note that the secondaries in post-CE binary central stars are often inflated with respect to that expected for a typical main sequence star of the same spectral type.  This leads to an increase in amplitude of the observed irradiation effect.}
\label{fig:reflection}
\includegraphics[width=\textwidth]{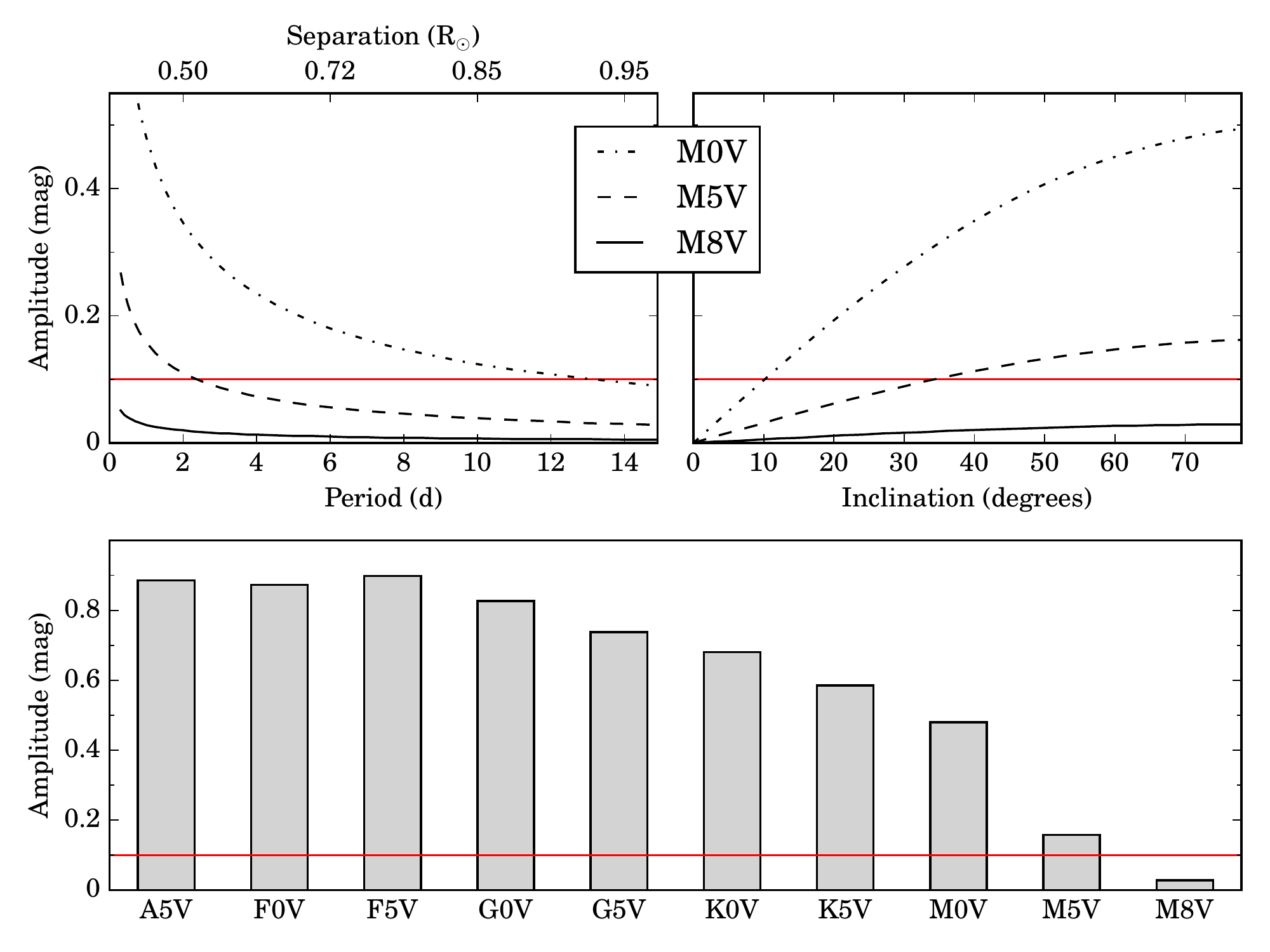}
\end{figure}

Beyond greatly increasing the sample of known binaries, the OGLE survey work also provided the least biased measure of the photometrically detectable binary fraction to date.  Biased only to systems with a central star observable by the survey (that is, brighter than $\sim$20 mag in the $i$-band and without significant nebular contamination), such stars present a photometrically-detectable close-binary fraction of $\sim$20\%.  It is important to stress that this number represents a lower limit for the true binary fraction.The smallest irradiation effect detected as part of the OGLE sample has an amplitude of $\sim$0.02 magnitudes. However, the associated system also shows eclipses that are approximately 0.5 magnitudes deep, thus making the variability much easier to detect.  Excluding these systems that display low-level irradiation effects but easily detectable eclipses, the smallest amplitude irradiation effects detected as part of the survey are $\sim$0.1 magnitudes.  The amplitude of an observed irradiation effect is a function of various parameters including the temperature and luminosity of the primary, temperature and luminosity of the secondary, and the binary period and inclination.  Figure \ref{fig:reflection} shows the amplitude of irradiation effect (as calculated using the state-of-the-art Wilson-Devinney code, PHOEBE \cite{prsa16}) for a hypothetical `base' system comprised of a hot 100,000 K central star (with other parameters based on evolutionary tracks \cite{millerbertolami16}) in a binary system of period 1d and an inclination of 70$^\circ$.  Each plot varies one parameter of the binary system, such as the period/separation, inclination or spectral type (and thus mass, temperature and luminosity) of the secondary.  
It is important to highlight that the secondaries of post-CE binary central stars are generally found to be inflated with respect to the radius expected for a typical main sequence star of the same spectral type \cite{jones15}.  This will act to increase the amplitude of the observed irradiation effect bringing the results for a given spectral type more in line with that expected for a star, perhaps, two or three spectral subtypes earlier (that is, the observed irradiation effect amplitude for an M8V secondary may, in fact, be more in line with those predicted for an M5V secondary).  It is clear that at short periods, high inclinations or for more massive secondaries, the 0.1 magnitude detection limit should result in a near-100\% completeness. However, for longer periods (beyond say 5--10 days) and, particularly, for less massive companions (even though the inflated secondaries are easier to detect), the completeness will drop dramatically \cite{demarco08}.  

Here, we have discussed the completeness of a survey, such as the OGLE survey, with a minimum detection amplitude of approximately 0.1 mag.  Targeted observations are capable of reaching much lower amplitudes (perhaps as low as 0.01 mag) at the cost of observing a much smaller sample. Particularly interesting in this respect is the discovery, using the Kepler satellite, of a short period, post-CE binary with variability at the level of 0.7 mmag (in this case, not due to an irradiation effect but rather a combination of doppler beaming and the tidal distortion of the two stars)€" -- well below the detection threshold from the ground \cite{demarco15}, even with targeted observations. More crucially, of the five central stars of PNe with usable data from Kepler, four showed variability (most below the threshold of what could be detected from the ground), and for three of them, this variability is possibly related to binarity (interestingly, none of the sample displayed an irradiation effect). This is obviously small number statistics, but it seems clear that the fraction of 20\% may have to be revised upwards. In any case, even the current fraction is too large to be explained by today's models, taking into account the binarity of solar-like stars and the tidal radii of giant stars on the red giant branch and on the AGB  \cite{mustill12,madappatt16}. If the above is true, this may mean that not only do binaries shape PNe, but they may also be a prerequisite for forming majority of them!
Furthermore, one should not forget that the CE phase may also lead to the merging of the two components  \cite{kochanek14}, thereby further increasing the fraction of {\it initial} close binary stars.
  
To these close binaries, one should also add the possibility of PNe harbouring wide binaries. 
A small group of PNe are known to harbour binary central stars in which a sub-giant or giant companion is enriched in carbon and slow-neutron-capture-process (s-process) elements \cite{bond03,miszalski12a,miszalski13a}. 
These PNe tend to present an apparent ring-like morphology, which is most likely the outcome of the mass transfer episode -- probably by wind -- that led to the pollution of the cool secondary star in carbon and s-process elements \cite{tyndall13}.
It is also interesting to note that long-term radial velocity monitoring has provided the first confirmation of orbital motion due to a long period binary in a PN, with the detections in BD+33$^\circ$2642, LoTr~5 and NGC1514 \cite{vanwinckel14,jones17b}.  Indeed, the period of NGC~1514 is so long that previous long-term ($\sim$ 1 year) monitoring attempts failed to recover any variability; extreme long-term monitoring over a period of nearly ten years was required to recover its periodicity, thus further highlighting the difficult in estimating the true binary fraction.

Other surveys that, in principle, should be sensitive to longer periods and less massive companions tend to report binary fractions much greater than 20\% but generally with smaller-number statistics and greater uncertainties \cite{demarco04,demarco13}.  For example, searching for cool companions based on their contribution to the spectral energy distribution of the central star spectrum (infrared excesses), and accounting for both detection limits and the possible contribution of white dwarf companions, results in a binary fraction of $\sim$80\% (with significant uncertainties), consistent with the observed fraction of aspherical PNe \cite{douchin15}. If this conclusion holds, this would imply that binarity is a near-necessity to form a planetary nebula \cite{moe06}.

\section*{Characterising the binaries and their host nebulae}
Great effort has been made to characterise the known population of binary central stars and their host nebulae.  For the binary central stars, this requires simultaneous modelling of light and radial velocity curves \cite{wilson71,prsa16,hillwig16b}.  In many cases, the intense irradiation effect leads to the production of emission lines in the `day-side' face of the secondary; in spite of the primary being much brighter than the secondary in almost all bands, radial velocity curves of both components can be derived (the primary's from the typical absorption lines of, e.g., He \textsc{ii}, N \textsc{v} and O \textsc{v}, and the secondary's from the emission lines C \textsc{iii}, C \textsc{iv} and N \textsc{iii}).  This leads to a model independent measurement of the mass ratio, which strongly constrains the later modelling.  Intriguingly, just as for UU Sagittae, main-sequence secondaries are found to be greatly inflated in all systems that have been subjected to detailed modelling \cite{jones15}.  This inflation is now generally considered to be an effect of mass transfer from the primary onto the secondary, either during the CE or, most likely, just prior to it.  The mass-transfer rate required to produce such high levels of inflation does need to be significant, but the phase may be short-lived such that the total mass transferred may be as low as a few hundredths of a solar mass \cite{prialnik85}.  This rapid mass transfer acts to knock the secondary out of thermal equilibrium, thereby causing the inflation.  The thermal timescale of the star is long enough that, even following the ejection of the CE and formation of the planetary nebula, the star remains inflated.  The high levels of irradiation may also contribute to maintaining this state, as borne out by the observed tendency of the secondary to display higher temperatures than isolated main sequence stars of the same mass \cite{demarco08}.

Further evidence of pre-CE mass transfer is provided in the central star system of The Necklace, where the main-sequence secondary is found to display an over-abundance of Carbon that is almost certainly due to carbon-rich material being accreted from the primary while it was on the AGB \cite{miszalski13b}.  Detailed spatio-kinematical modelling of the host nebulae also supports the idea that the binaries experience a period of mass transfer prior to the CE.  In many cases, the kinematical ages of jets or polar outflows in PNe with binary central stars are found to be older than the central regions of these nebulae \cite{jones14b}.  This can be understood as the jet being launched by the accretion disk during this episode of intense, pre-CE accretion, whereas the central region forms as a result of the ejection of the CE itself \cite{corradi11}. It is important to note that there are also planetary nebulae which show jets which are apparently younger than the central regions of their nebulae \cite{huggins07}, however only two such cases are known to host post-CE central stars.  These systems are somewhat harder to understand but it has been suggested that the jets may form from mass transfer from a secondary, which is overflowing its Roche lobe following the ejection of the CE, onto the primary (in the opposite direction compared to the pre-CE jet case) \cite{tocknell14,soker94}.  In proto-PNe, it has been suggested that an apparent correlation between the velocities of jets and equatorial tori may mean that they are formed by the same process \cite{huggins12}.  In any case, although there may be multiple mechanisms at work, it seems clear that binarity must play an important role in the formation of jets and tori.

Perhaps the most critical evidence for the importance of binarity in the shaping of PNe comes from combining the modelling of central binaries and their host nebulae.  Determining the true morphologies of PNe is a great challenge, given that one can only see these three-dimensional structures from a single angle and thus the assumed morphology is highly susceptible to projection effects \cite{kwok10}.  One way around this issue to to use kinematical information, through spatially-resolved spectroscopy, to recover the third-dimension \cite{garcia-diaz09,jones10b,jones12,huckvale13}.  Such spatiokinematical modelling has only been performed for a small number of PNe with binary central stars, but in each case where the nebular and binary inclinations have been derived, the binary orbital plane and symmetry axis of the nebula are found to lie perpendicularly.  Statistically speaking, the likelihood of finding such a correlation by chance is less than one in a million \cite{hillwig16}.  This provides clear evidence of the physical link between binarity and the nebulae, which is in keeping with all models of binary-driven shaping.

Efforts have been made to characterise the chemical abundances of PNe with known binary central stars and, just as for Abell 63, in all but one case the measured \emph{adf} was found to be greatly elevated.
In fact, the \emph{adf} of Abell~63 is on of the least-elevated amongst PNe with binary central stars, with others reaching up to to several hundred in their central regions \cite{liu06,corradi15,jones16a,wesson16}.  More recently, it has been shown that the high \emph{adf}s in these nebulae may result from the presence of two differently distributed gas phases: one of normal metallicity and temperature, and a second, more centrally-concentrated, low-temperature phase which is enhanced in metals \cite{garcia-rojas16,garcia-rojas16b,richer17}.  Such differing distributions may possibly be indicative of multiple episodes of mass loss and even fallback and reprocessing of ejected material, a hypothesis somewhat supported by the low masses measured in these objects (lower than expected for an entire AGB envelope ejected as part of a CE phase) \cite{corradi14,corradi15,jones16a}.

\section*{Another surprise}
\begin{figure}
\caption{Period distribution of known binary central stars with the companion type indicated where appropriate.  MS, Main sequence or giant companion; DD, double-degenerate systems where the companion is also a white dwarf; Unclassified, systems in which the companion type is uncertain or not well constrained by observations.}
\label{fig:periods}
\includegraphics[width=\textwidth]{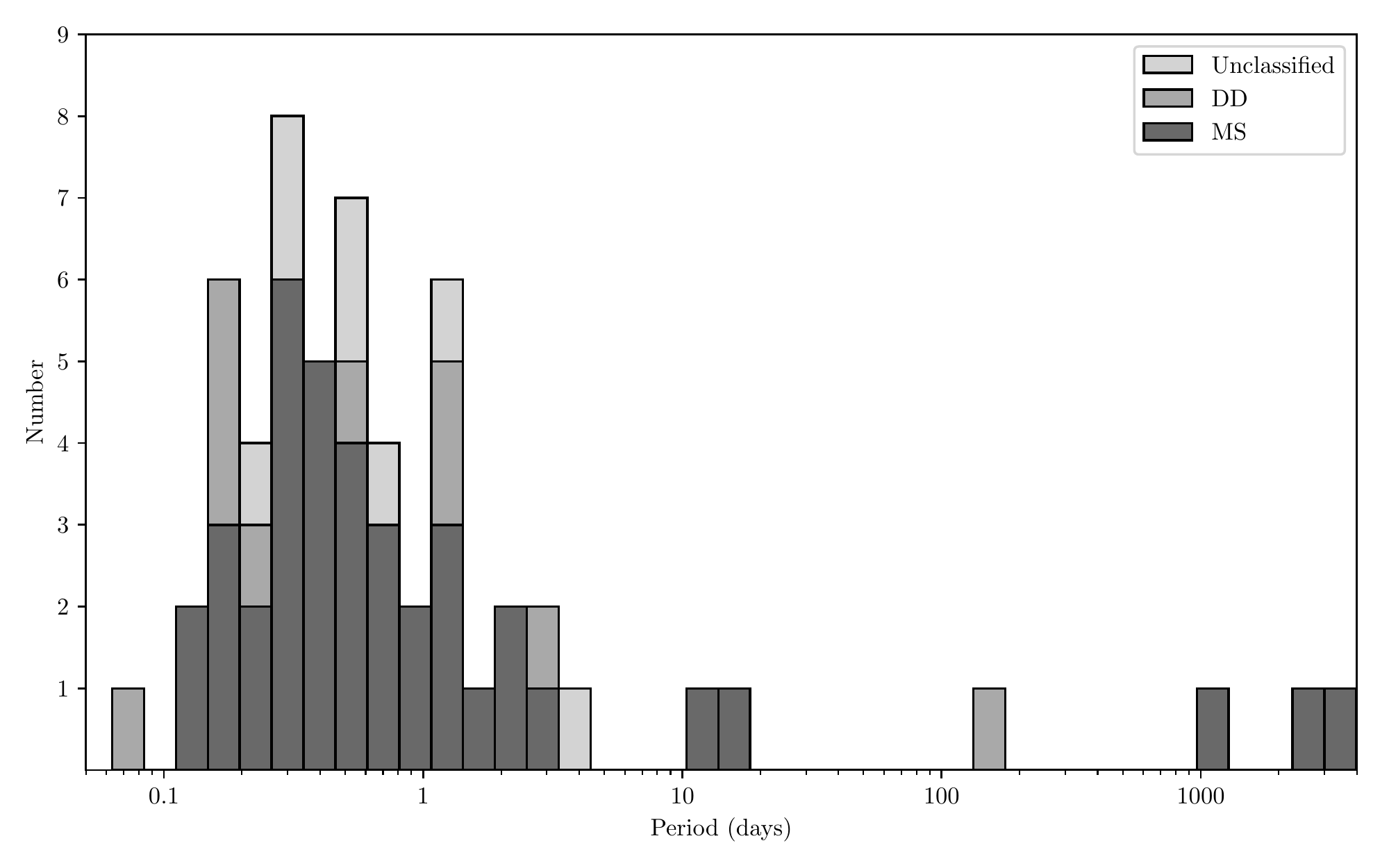}
\end{figure}

While most close-binary central stars comprise a low-mass main-sequence secondary, an appreciable number are systems where both components (primary and secondary) are post-AGB stars (also known as double-degenerate systems; \ref{fig:periods}).  These systems can be very difficult to detect as, unless the system is eclipsing or the orbital separation is very small, they do not display any photometric variability \cite{boffin12b}.  Given that these systems should be more difficult to detect, it is a surprise that around one fifth of all known binary central stars are double-degenerates, thus indicating that the true fraction should be much higher.
Such a high occurrence rate of double-degenerate systems is not predicted by common-envelope models  \cite{hillwig10,hillwig11}. Furthermore, the widest known binary central star in the post-CE domain (with a period of 142 days) is a suspected double-degenerate system, posing yet more problems for our understanding of the CE phase \cite{miszalski17}. These findings with respect to the formation of double degenerate systems have important consequences for the formation of type Ia supernovae, as one of the possible (and perhaps sole) formation channels is the merging of two white dwarfs, the total mass of which is greater than the Chandrasekhar limit \cite{tovmassian04}. In fact, the current best candidate for a type Ia supernova progenitor is the central star system of Hen~2-428, for which modelling has revealed the system to comprise twin 0.8 M$_\odot$ white dwarfs that are expected to merge in less than 700 million years \cite{santander-garcia15}.



\newpage

\section*{Acknowledgments}
This work makes use of data obtained from the Isaac Newton Group of Telescopes Archive which is maintained as part of the CASU Astronomical Data Centre at the Institute of Astronomy, Cambridge.  D.J. would like to thank Florencia Jim\'enez Luj\'an, P. Jones Jim\'enez and D. Jones Jim\'enez.

\noindent \textbf{How to cite this article:} Jones, D. \& Boffin, H. M. J. Binary stars as the key to understanding planetary nebulae. \textit{Nat. Astron.} 1, 0117 (2017).

\newpage

\end{document}